\documentclass[12pt]{article}
\usepackage{latexsym}
\usepackage{amsmath}
\usepackage{amssymb} 
\usepackage{relsize}
\usepackage{geometry}

\usepackage{tensor}
\usepackage{physics} %for Re Im
\usepackage{csquotes} % \textquote{ } -- for quotes

\usepackage{graphicx}       	% use this option to compile the pdf with images
\usepackage[font=small,labelfont=bf]{caption}
\usepackage{subcaption}

\def\beqn{\begin{eqnarray}}
\def\eeqn{\end{eqnarray}}

\def\beq{\begin{equation}}
\def\eeq{\end{equation}}
\def\ba{\beq\new\begin{array}{c}}
\def\ea{\end{array}\eeq}

\newcommand{\gsim}{\lower.7ex\hbox{$
\;\stackrel{\textstyle>}{\sim}\;$}}
\newcommand{\lsim}{\lower.7ex\hbox{$
\;\stackrel{\textstyle<}{\sim}\;$}}

\newcommand{\ntwo}{${\mathcal N}=2$ }

\newcommand{\ntwot}{${\mathcal N}= \left(2,2\right) $ }

\newcommand{\none}{${\mathcal N}=1$ }

\newcommand{\pt}{\partial}

\numberwithin{equation}{section}

\newcommand{\p}{\partial}

\newcommand{\ov}{\overline}

\def\slashed#1{\setbox0=\hbox{$#1$}             % set a box for #1
\dimen0=\wd0                                 % and get its size
\setbox1=\hbox{/} \dimen1=\wd1               % get size of /
\ifdim\dimen0>\dimen1                        % #1 is bigger
\rlap{\hbox to \dimen0{\hfil/\hfil}}      % so center / in box
#1                                        % and print #1
\else                                        % / is bigger
\rlap{\hbox to \dimen1{\hfil$#1$\hfil}}   % so center #1
/                                         % and print /
\fi}                                        %

\newcommand{\nbar}{\ov{n}}

\newcommand{\CP}{$\mathbb{CP}(N-1)~$}

\usepackage{mathtools}
\usepackage{hyperref}
\usepackage{xcolor}
\usepackage{tikz-cd}
\usepackage{placeins}
\usepackage[toc,page, title, titletoc]{appendix}
\begin{document}

\hypersetup{%
%colorlinks=false,% hyperlinks will be black
linkbordercolor=blue,% hyperlink borders will be =color
%pdfborderstyle={0 0 0.1}% 
}

%%%%%%%%%%%%%%%%%%%%%%%%%%%%%%%%%%%%%%%%%%%%%%%%%%%%%%%%%%%%%%%%%%%%%%%%%%%%%%%%%%
%
%	  		        T I T L E
%
%%%%%%%%%%%%%%%%%%%%%%%%%%%%%%%%%%%%%%%%%%%%%%%%%%%%%%%%%%%%%%%%%%%%%%%%%%%%%%%%%%

\begin{titlepage}

\begin{flushright}
FTPI-MINN-23-25

%December ??, 2023
\end{flushright} 

\begin{center}

% TITLE
{\Large{\bf 
% Renormalization and Effective Action in \textquote{Twisted-Mass} Deformed $\mathbb{CP}(N-1)$ Model
Large-$N$ Solution and Effective Action of \\[5pt] \textquote{Twisted-Mass} Deformed $\mathbb{CP}(N-1)$ Model
}}

\vspace{5mm}

{\large  \bf G.~Sumbatian$^{\,a,b}$, E.~Ievlev$^{\,c}$ and  A.~Yung$^{\,a,d}$}
\end{center}

% AFFILIATIONS
\begin{center}

$^{a}${\it National Research Center ``Kurchatov Institute'',
Petersburg Nuclear Physics Institute, Gatchina, St. Petersburg
188300, Russia}\\

$^{b}${\it  St. Petersburg State University, Universitetskaya nab., St.~Petersburg \\ 199034, Russia}\\

{\it  $^{c}$William I. Fine Theoretical Physics Institute,
University of Minnesota,
Minneapolis, MN 55455, USA}\\

{\it $^d$Higher School of Economics, National Research University, St. Petersburg \\
194100, Russia}

\end{center}

\vspace{5mm}

\begin{center}
{\large\bf Abstract}
\end{center}

We study effective dynamics of the non-supersymmetric two-dimen\-si\-o\-nal \mbox{$\mathbb{CP}(N-1)$} model in the large $N$ limit.
This model is deformed by a mass term $m$ preserving $\mathbb{Z}_N$ symmetry of the Lagrangian.
At small $m$ the theory is strongly coupled and resembles the undeformed \CP model, while at large $m$ it is in a weakly coupled Higgs phase with spontaneously broken $\mathbb{Z}_N$.
We find the phase transition point and discuss the fate of the kink-antikink \textquote{mesons} at strong coupling.
We also resolve an issue of instability that arose in previous studies of this model.

% keywords: large N, phase transition, deconfinement, marginal stability

\end{titlepage}

\setcounter{tocdepth}{2}
\numberwithin{equation}{section}
\tableofcontents

\clearpage

\section{Introduction}

Seiberg-Witten scenario of quark confinement, discovered in the seminal works \cite{SW1,SW2}, is based on a cascade gauge symmetry breaking.
The symmetry group surviving at low energies is Abelian, in the simplest case given by U(1).
A small mass $\mu$ for the adjoint multiplet breaks supersymmetry from \ntwo to \none and forces condensation of monopoles.
As a result, confining string of the Abrikosov-Nielsen-Olesen (ANO) type \cite{ANO} emerges in this theory.

Later this scenario was further extended to allow for a non-Abelian symmetry group at low energies.
Non-Abelian flux tubes (vortex strings) where originally discovered in the \ntwo supersymmetric QCD (SQCD) with gauge group U$(N)$ and $N_f=N$ flavors of quark hypermultiplets \cite{HT1,ABEKY,SYmon,HT2} (see \cite{Trev,Jrev,SYrev,Trev2} for a review). They   are responsible for the \textquote{instead-of-confinement} phenomenon in this theory \cite{SYdualrev}. 

In the \none SQCD limit (large $\mu$), the non-Abelian string ceases to be BPS saturated.
As a result, the low energy  theory on the non-Abelian string describing the dynamics of internal orientational moduli becomes non-supersymmetric, and turns out to be a two dimensional $\mu$-deformed sigma model with $\mathbb{CP}(N-1)$ target space \cite{YIevlevN=1,YIevlevN=1conference}.
(The deformation $\mu$ is a supersymmetry breaking parameter.)
Non-zero  quark masses in four dimensional SQCD yield a so-called twisted-mass \cite{Alvarez} deformation of this 
$\mathbb{CP}(N-1)$ model. This motivates the detail study of the dynamics of the mass deformed  non-supersymmetric \CP model.

In this paper we solve the twisted-mass deformed non-supersymmetric \CP model in the large $N$ limit.
Originally, the large $N$ approximation was used to solve non-supersymmetric as well as \ntwot \CP model at zero mass deformation \cite{W79,DAdda:1978vbw,Luscher:1978rn}.
In particular, it was shown that the auxiliary U(1) gauge field $A_\alpha$ becomes physical and remains massless in the non-supersymmetric case.
The low energy spectrum of this theory consists of neutral \textquote{mesons} formed by particle-antiparticle pairs of charged matter fields.
The $\mathbb{Z}_N$ symmetry remains unbroken.

The twisted mass deformation is introduced in the model with the help of another auxiliary complex scalar field $\sigma$ (in the supersymmetric case this field is actually a superpartner of the photon $A_\alpha$).
This scalar field, having dimension 1, naturally enters with a new dimensionless coupling constant.
In quantum theory this gives rise to a new dynamical scale $\Lambda_\sigma$ in addition to the usual scale $\Lambda$ of the \CP model  (both coupling constants are asymptotically free).
There is no reason to assume any particular relation between these two scales, and in this paper they are treated as independent.
In an earlier work \cite{Gorsky:2005ac}, however, a specific relationship between these two constants was assumed, which became a source of an instability (to be discussed below).

At small values of the deformation mass parameter $m$, the system stays in the strong coupling regime with unbroken $\mathbb{Z}_N$ symmetry. Namely, it is in the Coulomb/confinement phase found by Witten \cite{W79} in the  zero mass limit.
When $m$ is large enough, $\mathbb{Z}_N$ non-singlets develop vacuum expectation values (VEVs), breaking this symmetry. The theory now is in the weak coupling regime.

The large $N$ method employed in this paper allows us to identify these two phases and determine the phase transition point location.
Previously the existence of such a phase transition was argued in \cite{GSY05} based on flux tubes in a four-dimensional gauge theory and then this phase transition was studied in \cite{Gorsky:2005ac}. In this paper we revisit large-$N$ solution of 
\CP model  and resolve the issue of instability at small values of $m$ present in the solution found in \cite{Gorsky:2005ac}.

We were also able to show that the scalar field $\sigma$, auxiliary at the classical level, together with the photon becomes dynamical due to quantum effects at strong coupling.
Close to the phase transition point, $\sigma$ become light, so the kink-antikink mesons neutral under the unbroken global U(1)$^{N-1}$ decay. It is analogous to the curves of marginal stability in supersymmetric theories. At large $N$ the region where 
$\sigma$ becomes light stays finite, which allowed us to study this phenomenon.

Moreover $\sigma$ becomes massless
at the transition point and we  identify the conformal field theory at the critical point as a free theory of massless field
$\sigma$.

%The further phase transition happens in a narrow region which shrinks at large $N$, which makes our method insufficient for resolving it in details.
 
The paper is organized as follows.
In Sec.~\ref{sec:intro_phases} we introduce the \CP model and its twisted mass deformation and discuss the two phases in the weak coupling and strong coupling limits.
In Sec.~\ref{sec:vacuum_solution} we employ the large $N$ expansion to derive the effective potential and study vacua of this theory at different values of the mass deformation.
We also determine the phase transition point and calculate vacuum energies on both sides of the phase transition.
In Sec.~\ref{sec:effact} we derive the effective action and discuss dynamics in different phases in more details.
Sec.~\ref{sec:concl} presents our conclusions.
The Appendices contain some technical computational details.

\section{Overview of the \CP model and its phases}
\label{sec:intro_phases}

In this section we introduce the \CP model and it's twisted mass deformation $m$.
Then we briefly discuss two distinct phases that occur at small and large $m$ (in the sense to be defined below).

\subsection{\CP model action and its mass deformation}
\label{sec:the_model}

Let us start with the non-deformed model.
For our purposes, the most convenient representation of the \CP model action is the linear gauge sigma model formulation introduced in \cite{W79,DAdda:1978vbw}.
The auxiliary U(1) gauge field allows to cast the Lagrangian in a form bilinear in the complex matter fields $n^\ell$:
\begin{equation}
	S = \int d^2x \left\{|D_\alpha n^\ell|^2+\lambda\left(|n^\ell|^2- r_0\right)  \right\},
\label{Tone}
\end{equation}
where $D_\alpha = \partial_\alpha - i A_\alpha$. The matter field index runs $\ell = 1, \ldots, N$, and summation over repeated indices is assumed.
The field $\lambda$ is a Lagrange multiplier that enforces the \CP manifold condition,
\begin{equation}
	|n^\ell|^2 = r_0 \,.
\label{cpn_manifold_condition}
\end{equation}
The radius of the $\mathbb{CP}$ manifold $r_0$ plays the role of the (inverse) coupling constant of the theory.
By integrating out the auxiliary field $A_\alpha$ one can recover the non-linear sigma model Lagrangian.
This model enjoys SU($N$) global symmetry, with $n^\ell$ transforming in the fundamental representation.

This \CP model is asymptotically free \cite{P75}.
The bare coupling constant $r_0$ gives rise to the dynamical scale of the theory $\Lambda$ via the relation
\begin{equation}
	\Lambda^2 = M_{\rm uv}^2 \exp\left(-\frac{4\pi r_0}{N}\right)\,,
\label{Tseven}
\end{equation}
where $M_{\rm uv}$ is the UV cut-off scale.

\vspace{10pt}

The \textquote{twisted mass} deformation mentioned above is introduced by adding to the action \eqref{Tone} an extra term
\begin{equation}
	S_m = \int d^2x  \left\{  \sum_{\ell=1}^{N}|(\sigma-m_{\ell})n^\ell|^2 - \tau_0\sum_{\ell=1}^{N}|\sigma-m_{\ell}|^2\right\} \,.
\label{twisted_mass_deformation}
\end{equation}
Here, $\sigma$ is a new auxiliary complex scalar field.  Note that the masses in \eqref{twisted_mass_deformation} explicitly break the global SU$(N)$ symmetry present in \eqref{Tone},
\begin{equation}
    \mathrm{SU}(N) \to \mathrm{U}(1)^{N-1} \,.
\label{global_symmetry_breaking}
\end{equation}
We choose the masses as
\begin{equation}
	m_{\ell} = m\exp(\frac{2\pi i \ell}{N}),\quad \ell=0,1,...,N-1 \,,
\label{mass_circle}
\end{equation}
where the parameter $m$ is real and positive.
This choice of masses preserves $\mathbb{Z}_N$ symmetry:
\begin{equation}
	\sigma\rightarrow \exp(i\frac{2\pi k}{N})\sigma,\quad n_\ell\rightarrow n_{\ell+k} \,, \quad
	\text{for fixed } k = 1, \ldots, N \,.
\label{Z_N_symmetry}
\end{equation}
The first term in \eqref{twisted_mass_deformation} is motivated by supersymmetry \cite{Dorey} (the field $\sigma$ is actually a superpartner of $A_\alpha$ in the supersymmetric setup).
The second term in \eqref{twisted_mass_deformation} represents a \textquote{mass} term of the field $\sigma$.
Since $\sigma$ has classical dimension 1, the constant $\tau_0$ is actually dimensionless.
Note that $\tau_0$ enters the action \eqref{twisted_mass_deformation} with a sum of $N$ terms; in order to have a smooth large-$N$ limit, we need to hold 
\begin{equation}
	r_0 / N = \text{ fixed} \,, \quad
	\tau_0 = \text{ fixed} \,, \quad
	\text{ as } N \to \infty \,.
\end{equation}

In the context of the non-supersymmetric \CP model a similar mass deformation was studied in \cite{Gorsky:2005ac}. 
However, in the latter paper the second term in eq. \eqref{twisted_mass_deformation} was absent.

While on the classical level one could set $\tau_0$  to zero, on the quantum level this term is generated by loop corrections and therefore must be included. We will see later that this new coupling is also asymptotically free and 
 gives rise to its own dynamical scale $\Lambda_\sigma$ which can be defined by
\begin{equation}
	\Lambda_\sigma^2 = M_\text{uv}^2\exp(-4\pi \tau_0).
\label{Lambda_sigma_def}
\end{equation}
There is no a priori reason for assuming any particular relation between the scales $\Lambda$ and $\Lambda_\sigma$, so in this paper we treat these scales as independent.
Below we will see that the only constraint imposed by self-consistency is $\Lambda_\sigma > \Lambda$.

%It turns out that physical properties of our model depend on the parameter
%\begin{equation}
%	c \equiv \ln\frac{\Lambda_\sigma^2}{\Lambda^2} \,.
%\end{equation}
%
%This dependence will be studied below.

\subsection{Weak coupling regime}
\label{section_the_Higgs_phase}

Let us start with the classical analysis of the minima of the action \eqref{Tone} with the mass deformation \eqref{twisted_mass_deformation}.
Varying with respect to the Lagrange multiplier $\lambda$ gives the condition $|n^\ell|^2 = r_0$; 
one can always choose the vacuum configuration so that only one of the $n$-fields develops a non-zero VEV:
\begin{equation}
	n^{\ell_0} = \sqrt{r_0}, \quad \text{and } n^{\ell}=0 \text{ if } \ell\neq\ell_0.
\label{n_classconfig}
\end{equation}
Since a field charged under U(1) develops a VEV, and there is no dynamical massless gauge field in this phase, we call it a Higgs phase.

Varying the action \eqref{Tone} with respect to $\sigma$ gives an equation
\begin{equation}
	\sum_{\ell=1}^{N}(\sigma-m_{\ell})|n^{\ell}|^2-\tau_0\sum_{\ell=1}^{N}(\sigma-m_{\ell})	= 0 \,.
\label{sigma_classeq}
\end{equation}
Substituting \eqref{n_classconfig} into eq. \eqref{sigma_classeq}, the latter is solved by 
\begin{equation}
	\sigma = m_{\ell_0} \, \frac{r_0}{r_0-N\tau_0},  \qquad \ell_0=0,...,N-1 \,.
\label{classic_vacua_sigma}
\end{equation}
One can immediately see that at $r_0 = N \tau_0$ we ran into a trouble.
For now let us assume the inequality
\begin{equation}
	r_0 > N \tau_0 \,,
\label{couplings_inequality_classical}
\end{equation}
so that the VEV of $\sigma$ is finite.
Later in this paper we will see that a relation analogous to \eqref{couplings_inequality_classical} is actually necessary for self-consistency of this model on the quantum level.

The classical VEV of $\sigma$ in \eqref{classic_vacua_sigma} is proportional to the mass parameter $m$.
At large masses $m \gg \Lambda, \Lambda_\sigma$, the RG flow of the couplings is frozen at this large scale $\langle \sigma \rangle$.
The theory is at weak coupling, and as we will see below the vacuum configuration is still given in the large $N$ limit by eq. \eqref{n_classconfig} and \eqref{classic_vacua_sigma} with bare coupling constants replaced by  renormalized ones.
There are $N$ degenerate vacua labeled by the order parameter $\langle\sigma\rangle$; in each given vacuum the $\mathbb{Z}_N$ symmetry \eqref{Z_N_symmetry}   is spontaneously broken.

\subsection{Strong coupling regime} 

At small $m$ the theory flows to strong coupling, and the quasiclassical approximation is no longer applicable.
For $m=0$ this theory was solved at large $N$ in \cite{W79,DAdda:1978vbw,Luscher:1978rn}; let us briefly review here some features of this solution.

The VEVs of the $n^\ell$ fields at strong coupling vanish exactly, and SU($N$) symmetry of the theory is restored.
In the massive theory the $\mathbb{Z}_N$ symmetry \eqref{Z_N_symmetry} is  unbroken \cite{Gorsky:2005ac}.
The model has a unique ground state together with $\sim N$ quasi-stable local minima. 
The relative splittings between them is of order $\sim 1/N$, and false vacuum decay rates scale as $N^{-1}\exp (-N)$ \cite{nvacym,nvacymp} at large $N$.
The order parameter which distinguishes different quasivacua is the value of the constant
electric field or topological density
\begin{equation}
	Q=\frac{i}{2\pi}\,\varepsilon_{\alpha \beta}\,\pt^{\alpha} A^{\beta} =\frac1{8\pi r_0 } \,\varepsilon_{\alpha\beta}
		\,\pt^{\alpha}\bar{n}_i\pt^{\beta}n^i
\label{topdensity}
\end{equation}

At the one loop level in the large $N$ approximation, the (classically auxiliary) U(1) gauge field $A_\alpha$ acquires a kinetic term and thus becomes dynamical.
Kinks $n$ and anti-kinks $\bar{n}$, interpolating between the aforementioned quasi-vacua, have masses of order $\Lambda$ (the latter is defined by eq.~\eqref{Tseven}).
They are charged under the U(1) with a weak constant to mass ratio $e_\gamma / m_n \sim 1/\sqrt{N}$,
and transform as (anti-)fundamentals under the SU($N$).

The Coulomb potential in one-dimensional space is linear, which leads to formation of weakly bound states with zero net charge, $n\bar{n}$ mesons with constant electric field stretched between $n$ and $\bar{n}$ states.  There are no charged states in the spectrum.
This motivates  to call this Coulomb/confining phase. Above mentioned quasivacua are formed by   $n\bar{n}$ mesons,
so the $n$($\bar{n}$) state plays a role of a kink (antikink) separating region occupied by a meson from true vacuum,
see \cite{Gorsky:2005ac} for details.

%\EI{Add $\theta$}

\subsection{Transition point}

To reiterate, the behavior of \CP model is qualitatively different at small and large mass deformations.
\begin{itemize}
\item At small $m$ the ground state is unique, the $\mathbb{Z}_N$ symmetry as well as global U(1)$^{N-1}$ is unbroken, and the U(1) gauge field is massless. It is responsible for the confinement of charged $n$ states. We will call this phase the Coulomb/confining phase.
\item At large $m$ the ground state multiplicity is $N$, the $\mathbb{Z}_N$ symmetry is spontaneously broken, $n$ field develops VEV; we will call this phase the Higgs phase.
\end{itemize}
The order parameter that distinguishes between these two phases is the VEV of $\sigma$, or, equivalently, the VEVs of $n^\ell$ fields.

Below we are going to apply the large $N$ method to solve this model for all values of the mass deformation $m$. 
We will see the phase transition point and calculate the vacuum energy on both sides of this transition.

\section{Large-$N$ solution at non-zero mass deformation}
\label{sec:vacuum_solution}

In this section we use Witten's \cite{W79} large-$N$ method to solve the twisted-mass deformed \CP model.
The starting point for our calculations is the action comprised of \eqref{Tone} and \eqref{twisted_mass_deformation}:
\begin{equation}
	S = \int d^2x \left\{
		|D_\alpha n^\ell|^2 + \lambda\left(|n^\ell|^2- r_0\right)  
		+ \sum_{\ell=1}^{N}|(\sigma-m_{\ell})n^\ell|^2 - \tau_0\sum_{\ell=1}^{N}|\sigma-m_{\ell}|^2
	\right\}\,.
\label{class_action}
\end{equation}
Here, $D_\alpha = \p_\alpha - i A_\alpha$, masses are defined in \eqref{mass_circle}, while $r_0$ and $\tau_0$ are the bare couplings.

\subsection{Effective potential }

The first thing that we are interested in are the quantum vacua of the model under consideration.
These can be determined with the help of the effective potential.

In order to derive the  effective potential to the leading order at large $N$, we integrate over the $n^\ell$ fields in the path integral with the action \eqref{class_action}.
We are interested in Lorenz-invariant vacua; therefore, we will assume that the vector field $A_\alpha$ vanishes and that the scalar fields $\lambda$ and $\sigma$ are constant.
The action \eqref{class_action} is quadratic in $n_{\ell}$, so we are actually calculating a Gaussian integral.

As was outlined in Sec.~\ref{sec:intro_phases}, there are two phases depending on the value of $m$, with the order parameter being the VEVs of $n^\ell$.
Without loss of generality we can always take the field $n^\ell$ with $\ell = 0$ to have a non-zero expectation value in the Higgs phase.
% (all the other vacua can be obtained by an SU($N$) transformation).
Thus, in order to be able to distinguish between the Coulomb and the Higgs phases, we integrate only over the fields $n^\ell$, $\ell = 1, \ldots, N-1$, leaving the field $n^0 \equiv n$ as the order parameter.

Gaussian integration over the remaining $n^\ell$ yields a determinant
\begin{equation}
	\prod_{\ell=1}^{N-1}[\det(\partial_\alpha^2+\lambda+|\sigma-m_\ell|^2)]^{-1} \,.
\label{funcdet}
\end{equation}
%
%where we set $A_\alpha=0$ using Lorentz invariance of VEV. 
Calculation of this determinant is straightforward; introducing the UV cutoff $M_\text{uv}$ and keeping only the terms that depend on $\lambda$ and $\sigma$, we obtain the following contribution to the effective action:
\begin{equation}
	\frac{1}{4\pi}\sum_{\ell=1}^{N-1}(\lambda+|\sigma-m_\ell|^2)\left[\ln\frac{M_\text{uv}^2}{\lambda+|\sigma-m_\ell|^2}+1\right] \,.
\label{det_calculated}
\end{equation}
Now, from the definitions \eqref{Tseven} and \eqref{Lambda_sigma_def} we can express the bare couplings as
\begin{equation}
	r_0=\frac{N}{4\pi}\ln\frac{M_\text{uv}^2}{\Lambda^2} \,, \quad
	\tau_0=\frac{1}{4\pi}\ln\frac{M_\text{uv}^2}{\Lambda_{\sigma}^2} \,.
\label{r0}
\end{equation}
Adding together all the contributions to the effective potential and using \eqref{r0}, we finally arrive at
\begin{equation}
\begin{aligned}
	V_\text{eff}
		=& \left(\lambda+\left|\sigma-m_{0}\right|^{2}\right)|n|^{2} \\
		&+\frac{1}{4 \pi} \sum_{\ell=1}^{N-1}\left(\lambda+\left|\sigma-m_{\ell}\right|^{2}\right)\left[1-\ln 
			\frac{\lambda+\left|\sigma-m_{\ell}\right|^{2}}{\Lambda^{2}}\right] \\
		&+\frac{1}{4 \pi} \sum_{\ell=1}^{N-1}\left|\sigma-m_{\ell}\right|^{2} c \,.
\end{aligned}
\label{eff_potential_general}
\end{equation}
%
%\begin{equation}
%\begin{aligned}
%	V_\text{eff}
%		=& \lambda ( |n|^2 - r_\text{ren} + (N-1)/4\pi ) + \left|\sigma-m_{0}\right|^{2} |n|^{2} \\
%		&+ \frac{1}{4 \pi} \sum_{\ell=1}^{N-1} \left|\sigma-m_{\ell}\right|^{2} 
%			\left[  c + 1 - \ln\frac{\lambda+\left|\sigma-m_{\ell}\right|^{2}}{\Lambda^{2}}  \right] \\
%%		&+\frac{1}{4 \pi} \sum_{\ell=1}^{N-1}\left|\sigma-m_{\ell}\right|^{2} c \,.
%\end{aligned}
%\label{eff_potential_general}
%\end{equation}
%
The renormalized coupling $r_\text{ren}$ read off the effective potential is given by
\begin{equation}
	r_\text{ren} = \frac{1}{4\pi}\sum_{\ell=1}^{N-1}\ln\frac{\lambda+|\sigma-m_\ell|^2}{\Lambda^2} \,,
\label{rren}
\end{equation}
while constant $c$ parametrizes the relation between the two dynamical scale in this model,
\begin{equation}
	c \equiv \ln\frac{\Lambda_\sigma^2}{\Lambda^2} \,.
\label{c_def}
\end{equation}
Below we will see that for self-consistency we need $c > 0$.

Note that in \ntwot supersymmetric version of \CP model there is no need to introduce a second coupling constant $\tau_0$ due 
to the cancellation of the UV divergent term in \eqref{det_calculated} proportional to $\sum_{\ell}|\sigma-m_\ell|^2\ln {M_\text{uv}^2}$ with similar contribution coming from fermions, see \cite{SYhetN}. As a result there is only one scale $\Lambda$ in the supersymmetric version of the model.

By minimizing the effective potential \eqref{eff_potential_general} with respect to $\lambda$, $n$ and $\sigma$ to the leading order in large $N$, we obtain a set of equations that determine the vacua:
\begin{equation}
	|n|^2 = r_\text{ren} \,,
\label{vac_eq_1}
\end{equation}
\begin{equation}
	(\lambda+|\sigma-m_0|^2)n = 0 \,,
\label{vac_eq_2}
\end{equation}
\begin{equation}
	-\frac{1}{4 \pi} \sum_{\ell=1}^{N-1}\left(\sigma-m_{\ell}\right) \ln \frac{\lambda+\left|\sigma-m_{\ell}\right|^2}{\Lambda^2}
		+ \left(\sigma-m_0\right)|n|^2+\frac{N}{4 \pi} c \sigma
		= 0 \,,
\label{vac_eq_3}
\end{equation}
where $r_\text{ren}$ and $c$ are defined in \eqref{rren} and \eqref{c_def} respectively.
The vacua are found by solving these equations.

The first of them, eq. \eqref{vac_eq_1}, is a renormalized version of the bare condition \eqref{n_classconfig}.

The second equation \eqref{vac_eq_2} allows for two possibilities: either $n=0$ or $\lambda+|\sigma-m_0|^2=0$, $n \neq 0$.
The first option corresponds to the Coulomb/confining phase, while the second describes the Higgs vacuum.

The third equation \eqref{vac_eq_3} determines the VEVs of $\sigma$.
In \cite{Gorsky:2005ac} this equation was used to determine the value of $c$ in terms of the masses $m_\ell$, which actually lead to inconsistencies in the Coulomb/confining phase.
In this paper we rectify this by keeping $c$ as a fixed, independent of $m$ parameter defining the model .

Below we are going to investigate solutions to above equations. We will show that there is indeed a phase transition point at $m=\sqrt{c}\Lambda$, separating the Coulomb/confining and Higgs phases.

\subsection{The Higgs phase at large $N$}
\label{sec:higgs_solution}

\subsubsection{VEVs}
\label{sec:higgs_vevs}

At large $m$ we expect the VEVs of $n$ and $\sigma$ to flow to the classical values, eq. \eqref{n_classconfig} and \eqref{classic_vacua_sigma} respectively.
Since $n \neq 0$ in this limit, we expect that the Higgs phase is stretching from $m\to \infty$ down to the phase transition point.

The fact that $n \neq 0$ implies  that
\begin{equation}
	\lambda = -|\sigma-m_0|^2 \,,
\label{lambda_as_sigma_higgs}
\end{equation}
see  eq. \eqref{vac_eq_2}.

Substituting \eqref{lambda_as_sigma_higgs} and \eqref{vac_eq_1} into \eqref{vac_eq_3}, we obtain an equation on $\sigma$ only:
\begin{equation}
	- \frac{1}{4 \pi} \sum_{\ell=1}^{N-1} \left( m_0 - m_{\ell} \right) \ln \frac{\left|\sigma-m_{\ell}\right|^2-|\sigma-m_0|^2}{\Lambda^2}
		+\frac{N}{4 \pi} c \sigma
		= 0 \,.
\label{vac_eq_3_higgs}
\end{equation}
From the $\mathbb{Z}_N$ symmetry and matching with the classical solutions at large $N$ one might anticipate $N$ degenerate vacua and, correspondingly, $N$ different solutions for $\sigma$.
However, right from the start we have chosen not to integrate out the field $n^\ell$ with the specific index $\ell=0$, so that only $n^0$ can now develop a VEV. 
Therefore, eq.~\eqref{vac_eq_3_higgs} should actually give only one solution.
All the other $N-1$ vacua can be obtained by repeating this procedure but leaving $n^{\ell_0}$ with some other $\ell_0$ unintegrated over.

From the fact that $m_0 = m$ is real and positive and $\Re m_\ell \leqslant m$, one can show that the solution of eq.~\eqref{vac_eq_3_higgs}, $\sigma$, must be also real and positive.
With this knowledge we can simplify this equation; using the explicit formula for the masses \eqref{mass_circle}, after some algebra we arrive at (see Appendix~\ref{sec:sums} for the details):
\begin{equation}
%	m\ln\frac{\sigma}{\Lambda} + m\ln\frac{m}{\Lambda} + m - c\sigma = 0 \,.
	m \ln\frac{\sigma m}{\Lambda^2} + m - c\sigma = 0 \,.
\label{sigma_trans_eq}
\end{equation}

For $c>0$, eq.~\eqref{sigma_trans_eq} has exactly two solutions; these can be found approximately by the iteration method, see Appendix~\ref{sec:appendix_solving}.
At large $m$, the solution corresponding to the minimum of the effective potential is given by an approximate formula
\begin{equation}
	\sigma \approx \frac{m}{c}\ln\frac{m^2}{ c \Lambda^2} \,.
\label{largem_higgs_sigma}
\end{equation}
This gives the VEV of $\sigma$.
To obtain the VEV of $n$, we need to compute the renormalized coupling $r_\text{ren}$.
Substituting the expression for $\lambda$ eq.~\eqref{lambda_as_sigma_higgs} into the formula for $r_\text{ren}$ \eqref{rren}, we obtain after some algebra (see Appendix~\ref{sec:sums})
\begin{equation}
	r_\text{ren}(\sigma,m)
		= \frac{N}{4\pi}\ln\frac{\sigma m}{\Lambda^2} \,.
\label{rren_higgs_exact}
\end{equation}
This gives an exact expression for the renormalized coupling as a function of the field $\sigma$ and the parameter $m$ in the Higgs phase.
Using the approximate formula for the VEV of $\sigma$ \eqref{largem_higgs_sigma} we get an approximate formula for the VEV of $|n|^2$:
\begin{equation}
	|n|^2 = r_\text{ren}
		\approx \frac{N}{4\pi}\ln\frac{m^2}{c\Lambda^2} \,.
\label{largem_higgs_rren}
\end{equation}
As we see, at large $m$ it becomes large.

\subsubsection{Comparison with classical minimum}

It is instructive to compare the classical solution for $\sigma$, see eq.~\eqref{classic_vacua_sigma}, and the quantum VEV  given by eq.~\eqref{largem_higgs_sigma}.
Using definitions of the bare couplings eq.~\eqref{r0}, we can rewrite the classical formula \eqref{classic_vacua_sigma} for $\sigma$ as 
(recall that we are in the vacuum corresponding to $\ell_0=0$):
\begin{equation}
	\sigma_\text{class} 
		= m \frac{r_0}{r_0 - N\tau_0}
		= m \frac{r_0}{\frac{N}{4\pi}\ln\frac{M_\text{uv}^2}{\Lambda^2}-\frac{N}{4\pi}\ln\frac{M_\text{uv}^2}{\Lambda_{\sigma}^2}}
		= \frac{m}{c} \frac{4\pi}{N} r_0 \,.
\label{higgs_sigma_vev_compare_classical}
\end{equation}
On the other hand, using the quantum formulas eq.~\eqref{largem_higgs_sigma} and \eqref{largem_higgs_rren} we can express the VEV of $\sigma$ as
\begin{equation}
	\sigma_\text{quant}
		= \frac{m}{c} \frac{4\pi}{N} r_\text{ren} + \ldots \,,
\label{higgs_sigma_vev_compare_quantum}
\end{equation}
where the dots stand for terms subleading at large $m$.

Comparing eq.~\eqref{higgs_sigma_vev_compare_classical} and eq.~\eqref{higgs_sigma_vev_compare_quantum}, we see that they indeed agree to the leading order.
Namely, the quantum VEV of $\sigma$ can be obtained from the classical expression \eqref{higgs_sigma_vev_compare_classical} by replacing the bare coupling constant $r$ by the renormalized one.
At smaller $m$ quantum corrections to the classical formula start to become more pronounced.

Note that if coupling $\tau_0$ were equal to zero like in \cite{Gorsky:2005ac} we would have  $\sigma = m$ for the VEV of $\sigma$, see \eqref{higgs_sigma_vev_compare_classical}. Eq. \eqref{largem_higgs_sigma} shows that our result in this paper is logarithmically enhanced at large $m$.

\subsubsection{Vacuum energy and hints of a phase transition}

Vacuum energy is calculated as the value of the effective potential in its minimum determined by the vacuum equations.
First, let us compute the effective potential as a function of $\sigma$ only.
Substituting the expressions for 
$r_\text{ren}$ \eqref{rren} and
$\lambda$  \eqref{lambda_as_sigma_higgs}
into the effective potential \eqref{eff_potential_general}, 
after some algebra we arrive at (see Appendix~\ref{sec:sums} for details) 
\begin{equation}
	V_\text{eff}^\text{Hig} (\sigma,m)
		=-\frac{m^2}{4\pi}N \left[ 2\frac{\sigma}{m}  \ln\frac{ \sigma m }{\Lambda^2} -c\left(\left(\frac{\sigma}{m}\right)^2+1\right)\right] \,.
\label{higgs_potential}
\end{equation}
One can easily check that the extreme point condition $\p V_\text{eff}^\text{Hig} / \p \sigma = 0$ indeed gives the vacuum equation \eqref{sigma_trans_eq} above.

Note that the theory is stable only if $c>0$. Therefore, for self-consistency we require $c > 0$. This condition appears in the quantum theory  replacing the classical one  in \eqref{couplings_inequality_classical}.

At $m \gg \sqrt{c} \Lambda$ the minimizing value of $\sigma$ is given by eq.~\eqref{largem_higgs_sigma}.
The second derivative of the effective potential \eqref{higgs_potential} is positive at this point, which confirms that it is indeed a minimum of the potential.
Substituting \eqref{largem_higgs_sigma} into the effective potential \eqref{higgs_potential}, we obtain for the vacuum energy
\begin{equation}
	E^\text{Hig} \approx -\frac{ m^2  N }{ 4 \pi c } \left( \ln\frac{m^2}{ c \Lambda^2} \right)^2 \,.
\label{higgs_vac_energy_largem}
\end{equation}
Thus we see that at large $m$ vacuum energy becomes large and negative.

From the approximate formulas we can expect that the absolute value of vacuum energy \eqref{higgs_vac_energy_largem}, VEV of $\sigma$ \eqref{largem_higgs_sigma} and $r_\text{ren}$ \eqref{largem_higgs_rren}
approach small values, when we go towards smaller $m$.
However, since $r_\text{ren} = |n|^2$, we have a constraint
\begin{equation}
	r_\text{ren}\geq 0 \,.
\label{constraint}
\end{equation}
Therefore, $r_\text{ren}$ cannot decrease indeterminately.
As we will see shortly, $r_\text{ren}$ turns to zero at a finite value of $m$; this shows that a phase transition occurs at this value of $m$.

The phase transition point can be determined by solving simultaneously two equations, the vacuum equation and the vanishing of $r_\text{ren}$:
\begin{equation}
	\begin{cases}
		\pdv{ V_\text{eff}^\text{Hig} }{ \sigma } = 0  \\
		r_\text{ren}(\sigma, m) = 0
	\end{cases}
	\Longleftrightarrow
	\begin{cases}
		m \ln\frac{\sigma m}{\Lambda^2} + m - c\sigma = 0  \\
		\ln\frac{\sigma m}{\Lambda^2} = 0
	\end{cases}
\end{equation}
Here we used eq.~\eqref{sigma_trans_eq} and eq.~\eqref{rren_higgs_exact}.
The solution of this system is given by
\begin{equation}
	m_\text{crit} = \sqrt{c} \Lambda \,, \quad
	\sigma_\text{crit} = \frac{ \Lambda }{ \sqrt{c} } \,.
\label{transpoint_sigma}
\end{equation}

In fact this transition point also corresponds to another  phenomenon.
At large $m$ potential $V_\text{eff}^\text{Hig}$ in \eqref{higgs_potential} has two extrema in $\sigma$, one maximum and one minimum.
As we go towards smaller values of $m$, these two extrema get closer to each other.
Precisely at the point \eqref{transpoint_sigma} they merge; at this point not only the first but also the second derivative of the effective potential vanishes, 
\begin{equation}
	\pdv[2]{ V_\text{eff}^\text{Hig} }{ \sigma } \Bigg|_\text{crit} = 0 \,.
\label{higgs_second_derivative_vanish}
\end{equation}
At even smaller $m$ the potential $V_\text{eff}^\text{Hig}$ has no extrema anymore. 

Eq.~\eqref{higgs_second_derivative_vanish}
shows that the field $\sigma$ becomes massless at the transition point.  This suggests that
we are dealing with a phase transition of the second order and the field $\sigma$ is responsible for the critical behavior at the 
transition point. Below we will confirm these expectations.

\begin{figure}[h]
    \centering
    \begin{subfigure}[t]{0.49\textwidth}
        \centering
        \includegraphics[width=\textwidth]{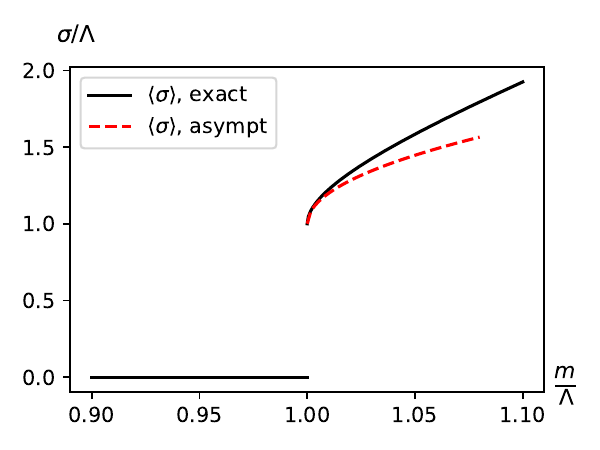}
        \caption{$\sigma$ is discontinuous}
        \label{fig:n_sigma_phasetrans_sigma}
    \end{subfigure}%
    ~ 
    \begin{subfigure}[t]{0.49\textwidth}
        \centering
        \includegraphics[width=\textwidth]{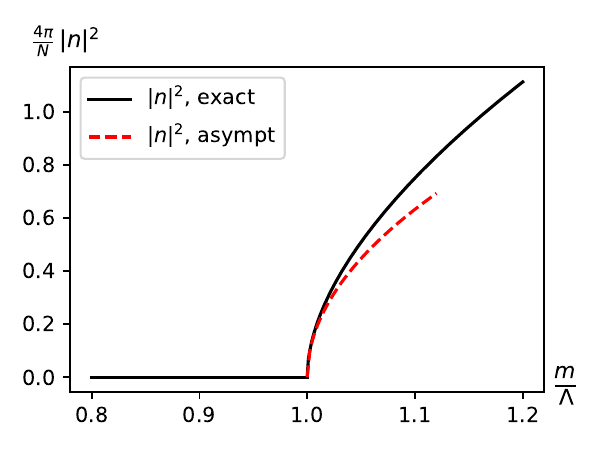}
        \caption{$|n|^2$ has a cusp}
        \label{fig:n_sigma_phasetrans_n}
    \end{subfigure}%
	\caption{
		Behavior of $\sigma$ and $|n|^2$ near the phase transition point $m = \sqrt{c}\Lambda$.
		(For this plot $c = 1$.)
		Black solid lines are the exact numerical solution, red dashed lines show the asymptotic expressions \eqref{sigmatr1} and \eqref{rren_near_phasetrans} respectively.		
	}
\label{fig:n_sigma_phasetrans}
\end{figure}

To conclude this subsection, let us analyze the behavior of the vacuum energy near the critical point \eqref{transpoint_sigma}.
We write $\sigma$ and $m$ as
\begin{equation}
	\sigma = \frac{ \Lambda }{ \sqrt{c} } + \delta \sigma \,, \quad
	m = \sqrt{c} \Lambda + \delta m \,,
\end{equation}
substitute this into the vacuum equation \eqref{sigma_trans_eq} and find an approximate%
\footnote{
	Note that because at the critical point \eqref{transpoint_sigma} the second derivative \eqref{higgs_second_derivative_vanish} vanishes, to find an approximate solution one needs to Taylor expand up to the third derivative in $\delta \sigma$, see Appendix~\ref{sec:appendix_solving}.
} %
solution:
\begin{equation}
	\delta\sigma \approx \frac{2}{c^{3/4}} \sqrt{ \Lambda \cdot \delta m } \,.
%	\delta\sigma \approx \frac{2\Lambda^{1/2}}{c^{3/4}} (\delta m)^{1/2} \,.
\label{sigmatr1}
\end{equation}
The behavior of the renormalized coupling and the VEV of $|n|^2$ is found by substituting eq.~\eqref{sigmatr1} into eq.~\eqref{rren_higgs_exact}:
\begin{equation}
	|n|^2 = r_\text{ren} \approx \frac{N}{2 \pi c^{1/4} }  \sqrt{ \frac{\delta m}{ \Lambda } } \,.
\label{rren_near_phasetrans}
\end{equation}
Note that $\sigma$ and $|n|^2 = r_\text{ren}$ as functions of $m$ have a cusp at the critical point \eqref{transpoint_sigma}.
For the visual representation see Fig.~\ref{fig:n_sigma_phasetrans}.

Substituting eq.~\eqref{sigmatr1} into eq.~\eqref{higgs_potential}, we get the vacuum energy near the transition point:
\begin{equation}
	E^\text{Hig} 
		\approx \frac{N}{4\pi}(c^2+1)\Lambda^2
		+ \frac{N}{2\pi}\frac{(c^2-1)\Lambda}{\sqrt{c}}\delta m
		+ O( (\delta m)^{3/2} ) \,.
			% +\frac{4N}{3\pi}\frac{\sqrt{\Lambda}}{c^{3/4}}(\delta m)^{3/2}+\frac{N}{4\pi}\frac{3c^2-23}{3c}(\delta m)^2.
\label{Ehiggs_near_phasetrans}
\end{equation}
This series runs in half-integer powers of $\delta m$.
Incidentally, the term linear in $\sim \sqrt{\delta m}$ is absent; this can be traced back to eq.~\eqref{higgs_second_derivative_vanish}.

\subsection{The Coulomb/confining phase at large $N$}

\begin{figure}[h]
	\centering
	\includegraphics[width=0.6\linewidth]{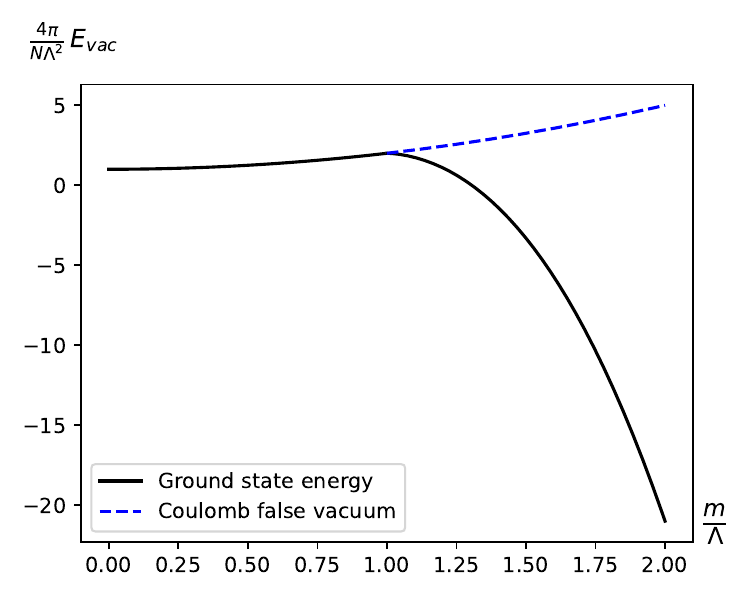}
	\caption{
		Vacuum energy and phase transition.
		Solid black line is the ground state (true vacuum) energy.
		Dashed blue line shows the would-be Coulomb phase energy analytically continued beyond the phase transition point $m = \sqrt{c}\Lambda$.
		(For this plot $c = 1$.)
	}
\label{fig:Evac}
\end{figure}

As we discussed above, the Witten's solution at zero $m$ \cite{W79} shows that $|n|^2$ vanishes in the quantum vacuum, see also \cite{DAdda:1978vbw,Luscher:1978rn}.
Moreover, as we just established, going down in $m$ from the Higgs phase one encounters a critical point at finite $m$ \eqref{transpoint_sigma} where $|n|^2=0$.

Therefore, it is reasonable to look for a solution of the vacuum equations where $|n|^2=0$ over a finite range of the mass parameter $m$.
The second vacuum equation \eqref{vac_eq_2} is then trivially satisfied, while the other two equations \eqref{vac_eq_1}, \eqref{vac_eq_3} are solved by
\begin{equation}
	\begin{cases}
		n = 0 \,, \\
		\lambda = \Lambda^2 - m^2  \,, \\
		\sigma = 0  \,.
	\end{cases}
\label{coulomb_vacs}
\end{equation}
Eq.~\eqref{coulomb_vacs} becomes an $m\neq 0$ generalization of the Witten's saddle point condition which was used to determine VEV of $\lambda$ in \cite{W79}, see also \cite{Gorsky:2005ac}.
Since the VEV of the $\sigma$ and $n$ field vanish, we are in a $\mathbb{Z}_N$ symmetric phase. 
Substituting \eqref{coulomb_vacs} into \eqref{eff_potential_general} we find vacuum energy in the Coulomb phase:
\begin{equation}
	E^\text{Clmb} = \frac{N}{4\pi} \left( \Lambda^2+m^2 c \right) \,.
\label{coulomb_vac_energy}
\end{equation}

To get more insight and to make sure that our solution \eqref{coulomb_vacs} indeed corresponds to a minimum, let us derive a formula for the effective potential as a function of $\sigma$.
In closed form this can be done in a small $\sigma$ approximation.
To integrate out $\lambda$, we write down the condition $r_\text{ren} = 0$ and Taylor expand around \eqref{coulomb_vacs}.
For a real $\sigma$ and $\lambda = (\Lambda^2 - m^2) + \delta\lambda$ we have
%
%\begin{equation}\label{sigma_lambda_behavior}
%\begin{aligned}
%&0=\frac{1}{4\pi}\sum\ln\frac{\lambda_0+\delta\lambda+|\sigma-m_{\ell}|^2}{\Lambda^2}\stackrel{\letus\sigma\in\Re}{=}\frac{1}{4\pi}\sum\ln\left[1+\frac{\delta\lambda+\sigma^2-2m_0\sigma\cos(2\pi\ell)}{\Lambda^2}\right]\approx\\
%&\approx\frac{1}{4\pi}\sum\left[\frac{\delta\lambda+\sigma^2-2m_0\sigma\cos(2\pi\ell)}{\Lambda^2}-\frac{(\delta\lambda+\sigma^2-2m_0\sigma\cos(2\pi\ell))^2}{2\Lambda^4}\right]\\
%&\approx\frac{N}{4\pi}\left(\frac{\delta\lambda+\sigma^2}{\Lambda^2}-\frac{m_0^2\sigma^2}{\Lambda^4}\right)
%	\Rightarrow\lambda=\lambda_0+\delta\lambda=\lambda_0+\sigma^2\left(\frac{m_0^2}{\Lambda^2}-1\right)
%\end{aligned}
%\end{equation}
%
\begin{equation}
	0 = r_\text{ren}
		= \frac{1}{4\pi} \sum\ln\frac{ (\Lambda^2 - m^2) + \delta\lambda + |\sigma-m_{\ell}|^2 }{\Lambda^2}
		\approx \frac{N}{4\pi}\left(\frac{\delta\lambda+\sigma^2}{\Lambda^2}-\frac{m^2\sigma^2}{\Lambda^4}\right) \,,
\end{equation}
from which obtain
\begin{equation}
	\lambda \approx \Lambda^2 - m^2 + \sigma^2 \left(\frac{m^2}{\Lambda^2}-1\right) \,.
\label{sigma_lambda_behavior}
\end{equation}
Substituting $|n|^2 = r_\text{ren} \equiv 0$ and $\lambda$ from eq.~\eqref{sigma_lambda_behavior} into the effective potential \eqref{eff_potential_general}, we obtain
\begin{equation}
	V_\text{eff}^\text{Clmb}
		\approx \frac{N}{4\pi} \left[ \Lambda^2 + c m^2 + c \sigma^2 \left(1-\frac{m^2}{c\Lambda^2}\right) \right] \,.
\label{coulomb_potential}
\end{equation}
From the potential in the form \eqref{coulomb_potential} we can immediately learn following consequences.

First of all, at $c > 0$ and $m < \sqrt{c} \Lambda$ the solution \eqref{coulomb_vacs} indeed corresponds to a stable theory, as second derivative of the effective potential is positive.
When $c < 0$, the effective potential does not have a minimum.
Thus for self-consistency we need to require $c > 0$; this agrees with our conclusions from the Higgs phase.

In this regard we remark that in \cite{Gorsky:2005ac}, $c$ was taken to be a certain function of the mass parameter $m$, which at small $m$ behaves as
\begin{equation}
	c \approx 1 + \ln\frac{m^2}{\Lambda^2} \,.
\label{constant_largeN}
\end{equation}
One can see that \eqref{constant_largeN} turns negative at sufficiently small $m$, which as we discussed leads to an instability.
We do not have this problem in the approach taken in the current paper. 

Next, at the critical point $m = \sqrt{c} \Lambda$ the second derivative of the potential \eqref{coulomb_potential} vanishes, and the field $\sigma$ becomes massless.
%This signifies an occurrence of an instability and a phase transition.
%Luckily, the critical value $m = \sqrt{c} \Lambda$ 
Luckily, this critical value of $m$ coincides with the critical value that we inferred from the Higgs phase, cf. \eqref{transpoint_sigma}. Thus, the field $\sigma$ becomes massless  when we move from both Coulomb/confining and Higgs phases 
to the phase transition point and is responsible for the critical behavior at this point. 
%These results further support the claim that the phase transition point is located at $m = \sqrt{c} \Lambda$.

One can actually show that we are dealing with a phase transition of the second order by comparing the vacuum energies in the Coulomb and Higgs phases on the two sides near the phase transition point $m = \sqrt{c} \Lambda$, eq.~\eqref{coulomb_vac_energy} and \eqref{Ehiggs_near_phasetrans} respectively.
The vacuum energy has the same value on both sides of this point,
\begin{equation}
	E^\text{Clmb} \Big|_\text{trans.pt.} = E^\text{Hig} \Big|_\text{trans.pt.} = \frac{N}{4\pi}(c^2+1)\Lambda^2 \,,
\label{Evac_trans_value}
\end{equation}
while the first derivative w.r.t. $m$ is discontinuous:
\begin{equation}
\begin{aligned}
	\frac{\partial}{\partial m} E^\text{Clmb} \Big|_\text{trans.pt.} 
		&= \frac{N}{2\pi}c^{\frac{3}{2}}\Lambda \,, \\[5pt]
	\frac{\partial}{\partial m} E^\text{Hig}    \Big|_\text{trans.pt.} 
		&=\frac{N}{2\pi}\left[c^{\frac{3}{2}}\Lambda-\frac{\Lambda}{\sqrt{c}}\right] \,.
\end{aligned}
\label{disc}
\end{equation}
This shows that the phase transition between Coulomb and Higgs phases is a second order phase transition. Moreover,
if we formally continue the Coulomb vacuum energy \eqref{coulomb_vac_energy} beyond the transition point $m=\sqrt{c}\,\Lambda$
to the region of larger masses it goes higher than the vacuum energy in the Higgs phase, see numerical results for the vacuum energy in Fig.~\ref{fig:Evac}. This confirms that $m=\sqrt{c}\,\Lambda$ is a transition point.

\section{More on dynamics in the  Coulomb/confining  and Higgs phases}
\label{sec:effact}

In this section we discuss the dynamics in the  Coulomb/confining  and Higgs phases in more details.

\begin{figure}[h]
    \centering
    \begin{subfigure}[t]{0.3\textwidth}
        \centering
        \includegraphics[width=\textwidth]{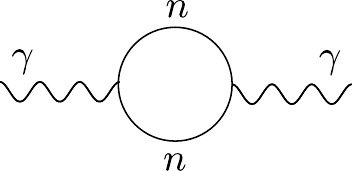}
        \caption{Photon kinetic term}
        \label{fig:feyndiagrams_photon}
    \end{subfigure}%
    \quad
    \begin{subfigure}[t]{0.3\textwidth}
        \centering
        \includegraphics[width=\textwidth]{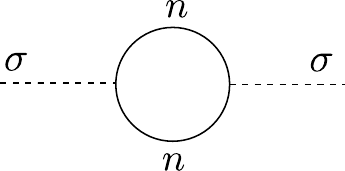}
        \caption{Scalar $\sigma$ kinetic term}
        \label{fig:feyndiagrams_sigma}
    \end{subfigure}%
	\caption{
		Loop corrections to the $A_\alpha$ and $\sigma$ propagators.
	}
\label{fig:feyndiagrams}
\end{figure}

\subsection{Effective action and light fields in the Coulomb phase}
\label{sec:effact_coulomb}

In this subsection we are going to derive and analyze the low-energy effective action in the  Coulomb/confining phase.

Witten's large-$N$ solution \cite{W79} demonstrates that at $m=0$ a kinetic term for the gauge field $A_\alpha$ is generated at one loop.
%This kinetic term is computed as a loop correction to the gauge field propagator.
In the twisted-mass deformed model considered here, there is another auxiliary field, $\sigma$, which also receives loop corrections.
The one-loop effective potential was computed in Sec.~\ref{sec:vacuum_solution}.
To find the full effective action, we need to compute the one-loop kinetic terms that arise in the theory at hand.

The corresponding loop diagrams are shown on Fig.~\ref{fig:feyndiagrams}; they were computed in \cite{Ievlev:2020qcy}.
The effective Lagrangian has the form
\begin{equation}
	\mathcal{L}^\text{Clmb} 
		= 
		%|D_\alpha n |^2 
		- \frac{1}{4 e_{\gamma}^2} F_{\alpha \beta}^2 
		+ \frac{1}{e_{\Re\sigma}^2}|\partial_\alpha \Re\sigma|^2 
		+ \frac{1}{e_{\Im\sigma}^2}|\partial_\alpha \Im\sigma|^2 
%		+ \frac{1}{ e_\lambda^2 } \lambda^2
%		- V^\text{Clmb}(\sigma) \,.
		- V_\text{eff} \,.
\label{coulomb}
\end{equation}
Here, $D_{\alpha}=\partial_{\alpha}-iA_{\alpha}$.
The last term is the effective potential \eqref{eff_potential_general}.
%As $\lambda$ remain non-dynamical, it can be integrated out; the resulting potential will  with $\lambda$ integrated out in the Coulomb phase; its expansion at small real $\sigma$ is given by eq.~\eqref{coulomb_potential}.
%
The inverse coupling constants in front of the kinetic terms in \eqref{coulomb} are given by 
\begin{equation}
\begin{aligned}
	\frac{1}{e_{\gamma}^2} &= \frac{1}{4 \pi} \sum_{\ell=1}^{N-1}\frac{1}{3} \frac{1}{\langle\lambda\rangle+|\langle\sigma\rangle-m_\ell|^2} \,, \\
	\frac{1}{e_{\Re\sigma}^2} &= \frac{1}{4\pi}\sum_{\ell=1}^{N-1}\frac{1}{3}\frac{(\langle\sigma\rangle- \Re m_\ell )^2}{(\langle\lambda\rangle+|\langle\sigma\rangle-m_\ell|^2)^2} \,, \\
	\frac{1}{e_{\Im\sigma}^2} &= \frac{1}{4\pi}\sum_{\ell=1}^{N-1}\frac{1}{3} \frac{( \Im m_\ell )^2}{(\langle\lambda\rangle+|\langle\sigma\rangle-m_\ell|^2)^2} \,.
%	\frac{1}{ e_\lambda^2 } &= \frac{1}{4 \pi} \sum_{\ell=1}^{N-1} \frac{1}{\langle\lambda\rangle+|\langle\sigma\rangle-m_\ell|^2} \,.
\end{aligned}
\label{coulomb_constants_general}
\end{equation}
with the VEVs $\langle \lambda \rangle$, $\langle \sigma \rangle$ determined from minimizing the effective potential, and $\langle \sigma \rangle$ is assumed to be real (this is a valid assumption as long as the parameter $m$ is real).
Substituting the Coulomb/confining phase VEVs from eq.~\eqref{coulomb_vacs}, we arrive at
%\begin{equation}
%	e_{\gamma}^2 = \frac{12\pi\Lambda^2}{N} \,, \quad
%	e_{\sigma}^2 = \frac{24\pi\Lambda^4}{N m^2} \,.
%\label{eren_coulomb}
%\end{equation}
%
\begin{equation}
	\frac{1}{e_{\gamma}^2 } = \frac{N}{12\pi\Lambda^2} \,, \quad
	\frac{1}{e_{\Re\sigma}^2 } = \frac{1}{e_{\Im\sigma}^2 } = \frac{N m^2}{24\pi\Lambda^4} \,. %\frac{1}{ e_\lambda^2 } = \frac{N}{ 4\pi\Lambda^2} \,.
\label{eren_coulomb}
\end{equation}
It is instructive to compare the result \eqref{eren_coulomb}, valid in the Coulomb/confining phase for $m$ in the interval $0 \leqslant m \leqslant m_\text{crit}$, to the Witten's result at $m=0$ \cite{W79}.
First, it turns out that $e_{\gamma}$ is exactly the same.
Second, at $m=0$ the scalar field couplings $1/e_{\Re\sigma}^2$ and $1/e_{\Im\sigma}^2$ vanish, and the field $\sigma$ remains
auxiliary without kinetic term while its VEV is given by \eqref{coulomb_vacs}; this again agrees with Witten's solution that did not have $\sigma$ at all.

Now, let us calculate masses of the fields entering the effective Lagrangian \eqref{coulomb}.
Since $n$ is the only field charged under U(1) and does not develop a VEV, the U(1) gauge field $A_\alpha$ is massless,
\begin{equation}
	m_\gamma = 0 \,.
\label{gamma_mass}
\end{equation}

Masses of   $n^\ell$ field are read off the action \eqref{class_action}.
In the Coulomb phase \eqref{coulomb_vacs} we have:
\begin{equation}
	m_{n^\ell}^2 = \langle\lambda\rangle + |\langle\sigma\rangle-m_\ell|^2
		= \Lambda^2 \,,
\label{quark_mass}
\end{equation}
which again coincides with Witten's result \cite{W79}.

However, as we discussed in Sec.~\ref{sec:intro_phases}, the spectrum actually consists of $n \nbar$ pairs. 
The binding energy $\Delta E$ is small at large $N$; we can estimate it as follows (see also \cite{W79} for an alternative derivation with the same result).
Given a separation $a$ between the two particles in the $n\nbar$ meson, the Coulomb potential due to the massless photon is given by
\begin{equation}
	U(a) = e_{\gamma}^2 a = \frac{12\pi\Lambda^2}{N} a \,.
\end{equation}
Let us take $a$ to be a characteristic distance between $n$ and $\nbar$ in the lowest energy state.
From the uncertainty principle we can then estimate the characteristic momentum of $n$ or $\nbar$ as $p \sim 1/a$, which makes the total binding energy of the meson
\begin{equation}
	\Delta E \sim \frac{p^2}{ m_n } + \frac{12\pi\Lambda^2}{N} a
		\sim \frac{ 1 }{ \Lambda a^2 } + \frac{12\pi\Lambda^2}{N} a \,.
\end{equation}
Minimizing this energy with respect to $a$ gives
\begin{equation}
	a \sim \frac{ N^{1/3} }{\Lambda} \,, \quad
	\Delta E \sim \frac{\Lambda}{ N^{2/3} } \,.
\end{equation}
These results confirm weak binding at large $N$.
The upshot of this is that we can estimate the mass of the lightest meson simply as twice the mass of an $n$ particle,
\begin{equation}
	m_\text{meson} = 2 m_n = 2 \Lambda \,.
\label{meson_mass}
\end{equation}

In order to correctly determine the mass of $\sigma$, we need to integrate out the non-dynamical field $\lambda$.
The corresponding potential $V_\text{eff}^\text{Clmb} (\sigma)$ for real $\sigma$ is given by eq.~\eqref{coulomb_potential}; the mass of $\Re\sigma$ is then calculated as
\begin{equation}
	m_{\Re\sigma}^2 = e_{\Re\sigma}^2 \cdot \frac{1}{2} \frac{\pt^2 V_\text{eff}^\text{Clmb} }{ \pt\sigma^2 } \Bigg|_{\sigma=0}
		= \frac{6c\Lambda^4}{ m^2} \left(1-\frac{m^2}{c\Lambda^2}\right) \,.
\label{sigma_mass_coulomb}
\end{equation}
From eq.~\eqref{eren_coulomb} and symmetry of the potential $V_\text{eff}$ \eqref{eff_potential_general} around $\sigma=0$, we also have $m_{\Im\sigma}^2 = m_{\Re\sigma}^2$.

%However, \eqref{sigma_mass_coulomb} makes sense in a low-energy effective theory only when $\sigma$ is a stable field.
At $m \to 0$, $\sigma$ becomes heavy and decouples, confirming our analysis above.
The only stable particles in the spectrum are the $n\nbar$ mesons with masses of $2 \Lambda$ \eqref{meson_mass}.
However, near the phase transition point $ m \lesssim \sqrt{c} \Lambda $, $\sigma$ becomes light, even lighter than the $n\nbar$ mesons.
The borderline value of the deformation parameter $m$ is determined by equating the scalar and meson masses:
\begin{equation}
	m_{\Re\sigma}^2 = m_\text{meson}^2 
	\Longrightarrow
	m =  \sqrt{ \frac{ 3 c }{ 5 } } \, \Lambda = \sqrt{ \frac{ 3  }{ 5 } } \, m_\text{crit} \,.
\label{newtp}
\end{equation}
At the value of the deformation parameter $m$ \eqref{newtp}, $\sigma$ and $n\nbar$-mesons are marginally stable.
Closer to the phase transition point those $n\nbar$-mesons that are singlets w.r.t. the unbroken U$(1)^{N-1}$ global symmetry \eqref{global_symmetry_breaking} decay into $\sigma$ or  pairs of $\sigma + \bar{\sigma}$, and the field $\sigma$ becomes absolutely stable.
At the phase transition point $m = \sqrt{c} \Lambda$ $\sigma$ becomes massless.

Thus, in the region of the deformation parameter
\begin{equation}
	\frac{3}{5}c\Lambda^2 < m^2 < c\Lambda^2
\end{equation}
the low-energy effective theory consists of a massless U(1) gauge field $A_\alpha$ and a scalar field $\sigma$. 
This is a free theory, since $\sigma$ does not interact with $A_\alpha$ in the leading order in $1/N$, and all the self-interactions are also suppressed at large $N$ (to see this one can rescale $\sigma \to \sigma/\sqrt{N}$).

Since in two dimensions the gauge field $A_\alpha$ does not have physical degrees of freedom we conclude that at the transition point
the conformal field theory, which describes the critical infrared dynamics is a theory of a free massless field $\sigma$.

%\EI{Estimate $n\nbar \to \sigma$ decay rates?}

\subsection{Discussion of the Higgs phase dynamics}

In the Higgs phase, $m > m_\text{crit}$, there are no massless photon, no confinement and no mesons. 
The spectrum consists of perturbations of the $n^\ell$ field and heavy kinks interpolating between $N$ vacua. Photon and $\sigma$ are heavier then 
lightest $n^\ell$'s  and therefore, do not represent stable bound states. To see this note that if, as in Sec.~\ref{sec:higgs_solution}, we are in the vacuum where $n^0 \neq 0$, then e.g. the mass of the $n^1$ field is suppressed at large $N$:
\begin{equation}
	m_{n^1}^2 = \langle\lambda\rangle + \left| \langle\sigma\rangle - m e^{ 2\pi i / N }  \right|^2
		= 2 m \langle\sigma\rangle \left( 1 - \cos\frac{2 \pi}{ N } \right)
		\sim \frac{1}{N^2} \,.
\end{equation}
Here we used the definition of masses eq.~\eqref{mass_circle}, expression for $\langle\lambda\rangle$ \eqref{lambda_as_sigma_higgs} and the fact that $\langle\sigma\rangle$ is real in the vacuum under consideration.

In principle, photon and $\sigma$ become massless at the transition point.  In the Higgs phase say, the photon mass squared
 is proportional to $r_\text{ren}$ and near the transition point where $r_\text{ren}=0$ behaves as
\beq
m_{\gamma}^2 \sim \Lambda^{3/2}\,\sqrt{\delta m},
\eeq
where we use \eqref{rren_near_phasetrans} and $\delta m =m -m_\text{crit}$. Therefore, photon becomes lighter then lightest 
of $n^\ell$'s at 
\beq
 \delta m\sim \frac{\Lambda}{N^4}.
\eeq

This is  too narrow interval 
of masses near the transition point and we have no control of this behavior in the large $N$ approximation, c.f.  
\cite{Gorsky:2005ac}. The same applies to the $\sigma$ field.

\section{Conclusions}
\label{sec:concl}

In this paper we studied the twisted-mass deformed \CP model at large $N$.
By calculating the effective potential and effective action in the leading order of $1/N$ expansion, we were able to study the vacua and the spectrum depending on the value of the mass deformation $m$.

Let us summarize the main results.
The model is asymptotically free and has two dynamical scales, $\Lambda$ corresponding to the \CP coupling constant, and $\Lambda_\sigma$ corresponding to the mass deformation.
It is convenient to describe the model in terms of one dimensionful parameter $\Lambda$ and one dimensionless parameter
\begin{equation}
	c \equiv \ln\frac{\Lambda_\sigma^2}{\Lambda^2} \,.
%\label{c_def}
\end{equation}
We found that this model is self-consistent only at $c > 0$.

Depending on the value of the deformation mass parameter $m$, we found two phases.
\begin{itemize}
\item 
At small $m$ the model is in the Coulomb/confining phase at strong coupling, resembling the Witten's solution \cite{W79}.
The \CP fields $n^\ell$ have zero VEVs and are charged under a dynamically generated massless photon $A_\alpha$.
There is also an extra massive complex scalar $\sigma$. The $\mathbb{Z}_N$ symmetry is unbroken.

\item
At large $m$ the model is in the Higgs phase at weak coupling.
Fields $n^\ell$ and $\sigma$ develop  VEVs. There are $N$ vacua and $\mathbb{Z}_N$ symmetry is broken. 

\end{itemize}
These phases are separated by a second order phase transition at $m_\text{crit} = \sqrt{c} \Lambda$. The conformal field theory which describes the infrared dynamics in the transition point is given by a free theory of massless field $\sigma$.

These results are consistent with the study of the \CP model with broken supersymmetry that arises as a world sheet effective theory in \none supersymmetric QCD \cite{YIevlevN=1,Gorsky:2019shz}.

At strong coupling and $ m < m_\text{crit} \sqrt{3/5} $ the spectrum consists of $n\nbar$-pair mesons with masses of $2 \Lambda$, bounded by  the linear Coulomb potential due to the massless photon.
However, the scalar $\sigma$ becomes light near the phase transition point; in the interval $ m_\text{crit} \sqrt{3/5} < m <  m_\text{crit}$ it becomes the lightest state in the spectrum, going massless 
at $m =  m_\text{crit} $.
Therefore, if we adiabatically raise $m$ end enter this interval, the $n\nbar$ mesons that are singlets w.r.t. the U$(1)^{N-1}$ global symmetry \eqref{global_symmetry_breaking} decay, forming $\sigma$ particles.
This picture can also be reversed: near the phase transition point $\sigma$ is  stable, but decays to $n\nbar$ kink-antikink mesons for lower values of $m$.

Above the phase transition point in the Higgs phase, $m > m_\text{crit}$, there are no massless photon and  no confinement. 
The spectrum consists of weakly interacting  $n^\ell$ fields and heavy solitons. Photon and $\sigma$ are heavier then 
lightest $n^\ell$'s  and  do not represent stable bound states. 

We can draw parallels with the instead-of-confinement phase in asymptotically free versions of \ntwo supersymmetric QCD SQCD \cite{SYdual}, see also \cite{SYdualrev} for a review.
These theories support non-Abelian strings responsible for confinement of monopoles.
The low energy theory on the world sheet of such a string is a supersymmetric (weighted) \CP sigma model.

In \cite{Ievlev:2020qch,Ievlev:2021mcm} the instead-of-confinement phenomenon was explicitly demonstrated: perturbative states that exist at weak coupling (corresponding to screened quarks and Higgsed gauge bosons), decay into confined monopole-antimonopole pairs at strong coupling.
A similar thing happens in the current model at strong coupling: the field $\sigma$ decays into $n\nbar$ mesons when we lower the parameter $m$.

\vspace{10pt}

%\section*{List of needed figures}
%
%
%
%Make figures:
%\begin{enumerate}
%\item Higgs: evolution of extrema of Veff near phase trans.
%\end{enumerate}
%
%Higgs phase:
%
%\begin{figure}[h]
%\center
%\includegraphics[width=0.75\linewidth]{r_ren_many.jpg}
%\caption{ Dependence of $\frac{4\pi}{N}r_\text{ren}$ on parameter $\frac{m}{\sqrt{c}\Lambda}$. $r_\text{ren}$ tends to 0 as $m\rightarrow\sqrt{c}\Lambda$. }
%\label{fig5}
%\end{figure}
%
%\begin{figure}[h]
%\center
%\includegraphics[width=0.75\linewidth]{sigma_higgs.jpg}
%\caption{$\sigma$ as function of $m_0$ for different $c$.}
%\label{fig3}
%\end{figure} 
%
%\begin{figure}
%\center
%\includegraphics[width=0.75\linewidth]{Evac_higgs.jpg}
%\caption{$E_{vac}^{(Higgs)}$ as function of $m_0$ for different $c$.}
%\label{fig4}
%\end{figure}
%
%Strong phase:
%
%\begin{figure}[h!]
%\center
%\includegraphics[width=0.75\linewidth]{Evac_HandC.jpg}
%\caption{Vacuum energies versus $m$ with $c,\Lambda=1$. The orange line is the Higgs vacuum energy, the blue line for the Coulomb/confining one and has the formal extrapolation to unphysical values of $m$ above the transition point.  }
%\label{fig1}
%\end{figure} 
%
%
%
%\clearpage

\section*{Acknowledgments}

The authors are grateful to  M. Shifman for valuable discussions. 
% The work of A.Y. was  supported by Skolkovo Institute of Science and Technology.
This work is supported in part by the Foundation for the Advancement of Theoretical Physics and Mathematics \textquote{BASIS},  Grant No. 22-1-1-16.

\appendix

\section{Technicalities of large-$N$ summations}
\label{sec:sums}

Throughout this paper we repeatedly dealt with finite sums in the large-$N$ limit.
These sums can be calculated by passing to the Riemann definite integral.
In this Appendix we present the details of these calculations.

We are going to need the following formulas:
\begin{equation}
	\sum_{\ell=1}^{N-1}\left[1-\cos\left(2\pi\frac{\ell}{N} \right)\right] = N \,,
\label{sum_1}
\end{equation}
\vspace{10pt}
\begin{equation}
\begin{aligned}
	\sum_{\ell=1}^{N-1} &\ln\left[ 1-\cos\left(2\pi\frac{\ell}{ N} \right)\right]
		= N\sum_{\ell=1}^{N-1}\ln\left[ 1-\cos\left(2\pi\frac{\ell}{ N} \right)\right] \frac{1}{N} \\
		&\stackrel{N\rightarrow\infty}{=} N\int_0^1\ln(1-\cos(2\pi x))dx
		=-N\ln 2 \,,
\end{aligned}
\label{sum_2}
\end{equation}
\vspace{10pt}
\begin{equation}
\begin{aligned}
	\sum_{\ell=1}^{N-1} & \left(1-\cos\left(2\pi\frac{\ell}{ N} \right)\right) 
				\ln\left[1-\cos\left( 2\pi\frac{\ell}{ N} \right)\right] \\
%		&= N\sum_{\ell=1}^{N-1}\left(1-\cos\left(2\pi\frac{\ell}{ N} \right)\right) 
%				\ln\left[ 1-\cos\left(2\pi\frac{\ell}{ N} \right)\right] \frac{1}{N} \\
		&\stackrel{N\rightarrow\infty}{=} N\int_0^1(1-\cos(2\pi x))\ln(1-\cos(2\pi x))dx
		=N(1-\ln2) \,.
\end{aligned}
\label{sum_3}
\end{equation}

With these results at hand, we can calculate what we need.
Let us start by computing $r_\text{ren}$ in the Higgs phase, for concreteness choosing a vacuum where $\sigma$ is real.
Using eq.~\eqref{mass_circle} and eq.~\eqref{lambda_as_sigma_higgs} in eq.~\eqref{rren} we see that it involves the sum of the terms of the form
\begin{equation}
\begin{aligned}
	\ln\frac{\left|\sigma-m_{\ell}\right|^2-|\sigma-m_0|^2}{\Lambda^2}
		&= \ln\frac{2\sigma(m_0-\Re m_{\ell})}{\Lambda^2} \\
		&= \ln\frac{2\sigma m}{\Lambda^2} + \ln\left[1-\cos\left(2\pi\frac{\ell}{N}\right)\right]
\end{aligned}
\label{OUF1}
\end{equation}
Carrying out the summation $\frac{1}{4\pi} \sum_{\ell=1}^{N-1} $ and using eq.~\eqref{sum_2} we obtain
\begin{equation}
	r_\text{ren} = \frac{ N }{ 4 \pi } \ln\frac{2\sigma m}{\Lambda^2} - \frac{ N }{ 4 \pi } \ln 2 = \frac{ N }{ 4 \pi } \ln\frac{\sigma m}{\Lambda^2}
\end{equation}
Thus we prove eq.~\eqref{rren_higgs_exact}.

The vacuum equation \eqref{sigma_trans_eq} for real $\sigma$ is derived similarly.
Using eq.~\eqref{mass_circle} and eq.~\eqref{lambda_as_sigma_higgs}, we rewrite each term of the sum in eq.~\eqref{vac_eq_3_higgs} as
\begin{equation}
\begin{aligned}
	\left( m_0 - m_{\ell} \right) & \ln \frac{\left|\sigma-m_{\ell}\right|^2-|\sigma-m_0|^2}{\Lambda^2} \\
		&= m \left[1-\cos\left(2\pi\frac{\ell}{N} \right)\right] 
			\left( \ln\frac{2\sigma m}{\Lambda^2} + \ln\left[1-\cos\left(2\pi\frac{\ell}{N}\right)\right] \right)
\end{aligned}
\end{equation}
This can be easily summed over $\ell$ with the help of eq.~\eqref{sum_1} and eq.~\eqref{sum_3}.
Assembling all the contributions we recover eq.~\eqref{sigma_trans_eq}.

To derive the effective potential in the form eq.~\eqref{higgs_potential}, we take the original formula \eqref{eff_potential_general} and rewrite (again, assuming $\sigma$ to be real)
\begin{equation}
	\abs{ \sigma - m_\ell }^2 = (\sigma - m)^2 +2 \sigma m \left[ 1 - \cos\left(2\pi\frac{\ell}{N} \right) \right] \,.
\end{equation}
After that, the large-$N$ sums are calculated as above, with the resulting eq.~\eqref{higgs_potential}.

\section{Details of solving vacuum equations in the Higgs phase}
\label{sec:appendix_solving}

First, let us comment on the iteration method of solving eq.~\eqref{sigma_trans_eq},
\begin{equation}
	m \ln\frac{\sigma m}{\Lambda^2} + m - c\sigma = 0 \,.
\label{sigma_trans_eq_appendix}
\end{equation}
This equation has two solutions at $c > 0$.

To find the solution corresponding to \eqref{largem_higgs_sigma} at large $m$ and $c > 0$, we rewrite eq.~\eqref{sigma_trans_eq_appendix} as
\begin{equation}
	\sigma = \frac{m}{c}\left[\ln\left(\frac{\sigma m}{\Lambda^2}\right)+1\right] \,.
\label{iter_1}
\end{equation}
Taking $\sigma^{(0)} = m / c$ as the starting approximation, we get for the first iteration
\begin{equation}
	\sigma^{(1)} = \frac{m}{c}\left[\ln\left(\frac{m^2}{ c \Lambda^2}\right) + 1  \right] \,,
\end{equation}
which coincides with \eqref{largem_higgs_sigma} at large $m$.
Note that at large $m$, the derivative of the r.h.s. of eq.~\eqref{iter_1} is $\sim m / \sigma \sim 1/ \ln m \ll 1$, which shows that it is a contraction mapping, thus validating this iteration scheme.

The second solution of eq.~\eqref{sigma_trans_eq_appendix} actually corresponds to a maximum.
It can be found by rewriting eq.~\eqref{sigma_trans_eq_appendix} in the form
\begin{equation}
	\sigma = \frac{ \Lambda^2 }{ m } \, e^{c \sigma / m - 1 }
\label{iter_2}
\end{equation}
and performing iterations. 
For the starting approximation it is suggestive to take $\sigma^{(0)} = 0$, then the first iteration gives
\begin{equation}
	\sigma^{(1)} = \frac{\Lambda^2}{me} \,.
\end{equation}
This iteration procedure is valid for any value of $c$.
One can show that for large enough $m$ the map \eqref{iter_2} is also contracting.

\vspace{10pt}

Lastly, let us comment on solving the vacuum equations near the phase transition point.
Take the effective potential $V(\sigma, m)$ with $\lambda$ and $n$ integrated out, then the vacuum equation can be written as
\begin{equation}
	V' (\sigma, m) = 0 \,,
\label{vac_eq_appendix}
\end{equation}
where the prime denotes a derivative w.r.t. $\sigma$.
Now, near the phase transition point at $m = m_\text{crit}$ we can do perturbation theory on eq.~\eqref{vac_eq_appendix}.
Writing $\sigma = \sigma_\text{crit} + \delta\sigma$, $m = m_\text{crit} + \delta m$ and recalling that at the phase transition point $V'' = 0$ (see eq.~\eqref{higgs_second_derivative_vanish}), we have to the lowest non-trivial order:
\begin{equation}
	\frac{ V''' }{2}  \, ( \delta\sigma )^2 + \pdv{ V' }{m} \delta m \approx 0 \,,
\end{equation}
from which we find, writing the derivatives explicitly,
\begin{equation}
	\delta \sigma \approx (\delta m )^{1/2} \cdot \sqrt{ -2  \pdv{V}{m}{\sigma} \Bigg/ \pdv[3]{V}{\sigma} }  \ .
\end{equation}
The derivatives here are calculated at the phase transition point \eqref{transpoint_sigma}.
Evaluating this expression with the potential \eqref{higgs_potential} yields the solution \eqref{sigmatr1}.

%%%%%%%%%%%%%%%литература%%%%%%%%%%%%%%%%
\clearpage

%\section{REMOVE --- BIBLIOGRAPHY STARTS HERE}

%\bibitem{elf},
%A.~P.~Prudnikov, Yu.~A.~Brychkov, O.~I.~Marichev, N.~M.~Queen,
%{\em \textbf{INTEGRALS AND SERIES}, vol. 1. \textbf{Elementary Functions}} --- Sec. 4.4.6, 4.4.7 ---
%ISBN 2-88124-089-5.
%% page 644(648)

\end{document}